\documentclass[acmsmall]{acmart}

\AtBeginDocument{%
  }

\setcopyright{cc}
\setcctype{by}
\acmJournal{PACMNET}
\acmYear{2026} \acmVolume{4} \acmNumber{CoNEXT4} \acmArticle{44}
\acmMonth{12} \acmDOI{10.1145/3856955}

\usepackage{subcaption}

\usepackage{booktabs}
\usepackage{pifont}
\newcommand{\cmark}{\ding{51}}
\newcommand{\xmark}{\ding{55}}

\begin{document}

\title{A Probabilistic Optical-Layer Surrogate for Risk-Aware Control-Plane Decision Making}

\begin{abstract}
In this paper, we propose a generative surrogate for the optical layer. Current control-plane decisions in optical transport networks are made using analytical impairment models, deterministic learned predictors, or quantile-regression-based uncertainty estimates. These approaches expose only a limited, static summary of the optical layer at any given time. This forces the control plane to rely on conservative fixed margins that leave capacity underutilized. Our proposed surrogate is a conditional generative model trained on traces from optical simulators with validated physical models. The surrogate produces a full conditional distribution over physical-layer metrics such as OSNR, Q-margin, nonlinear penalty, and feasibility. This output allows the controller to estimate violation probability and tail risk at arbitrary thresholds at runtime. This is different from existing approaches because no retraining is required for each decision. We present the first case study in fast path feasibility estimation using a generative surrogate for the optical layer. We compare the surrogate against simulators, deterministic learned predictors, and quantile-regression estimators. We evaluate the surrogate along nine research questions (RQ1--RQ9) covering prediction accuracy, inference latency, uncertainty quality, controller integration, cross-emulator generalization, dynamic reconfiguration under span degradation, topology scalability, GNPy-validated generalization, and finally, inference latency on real-world live hardware traces. The core contribution is an optical-layer control abstraction where physical-layer behavior is exposed as a fast probabilistic query mechanism instead of an offline simulation output or a deterministic feasibility rule.
\end{abstract}



\received{5 June 2026}
\received[accepted]{10 September 2026}

\author{Venkat Sai Suman Lamba Karanam}
\affiliation{%
  \department{Computer Science}
  \institution{Bowling Green State University}
  \city{Bowling Green}
  \state{Ohio}
  \country{USA}}
\email{vlambak@bgsu.edu}
\maketitle

\section{Introduction}

Current optical network control plane decisions are routinely made based on physics-based simulators that model the underlying optical-layer behavior~\cite{rafique2018mlautomation,pointurier2021mlqot,sartzetakis2019accurate}. Three broad classes of state-of-the-art approaches exist as follows. (1) Analytical impairment models based on the Gaussian Noise (GN) and enhanced GN (EGN) models~\cite{poggiolini2012gn} provide point predictions on closed-form OSNR and Q-margin metrics, but these predictions are made based on a static commissioning snapshot of the network, from which accurate per-span physical parameters are computed. (2) Physics-based simulators such as GNPy~\cite{gnpy} are the de facto ground truth for control plane decisions in the current standards. While these approaches provide higher fidelity predictions than analytical impairment models, they are expensive offline tools, both latency-wise and computationally. Although simulators can be rerun when the network state changes, the latency and computational cost for each simulator run makes it impractical at the control plane on a per-decision basis. (3) Deterministic predictors~\cite{pointurier2021mlqot,sartzetakis2019accurate} train regression models on simulator-generated traces to make point predictions that are used at the control plane. These still carry the limitations from point estimations and remain tied to the static configurations used to train the regression models. While quantile regression and conformal prediction methods~\cite{koenker1978regression,koenker2001quantile,mahajan2021retraining} provide certainty bounds instead of point estimates to overcome this limitation -- however, these still produce only a fixed, pre-specified summary of the distribution i.e., partial distribution (e.g., P10, P50, P90) rather than the full -- specifically, the conditional distribution over physical-layer metrics like OSNR and Q-margin given a candidate optical path and its configuration i.e., $p(\text{OSNR}, \text{Q-margin} \mid x)$, which would let the controller query arbitrary violation probabilities at runtime. Across all the existing approaches, physical-layer behavior is exposed only as a static summary. This setting forces control plane operators to extend budget margins (e.g., for spectrum allocation) so that the unrepresented uncertainty not captured by the static snapshot does not leave capacity underutilized, meets the QoS, does not violate quality-of-service requirements, and does not result in unnecessary connection blocking (increased blocking probability).
\begin{figure}[htp]
\centering
    \includegraphics[width=0.75\linewidth,trim={4.75cm 6cm 4.0cm 2.25cm},clip]{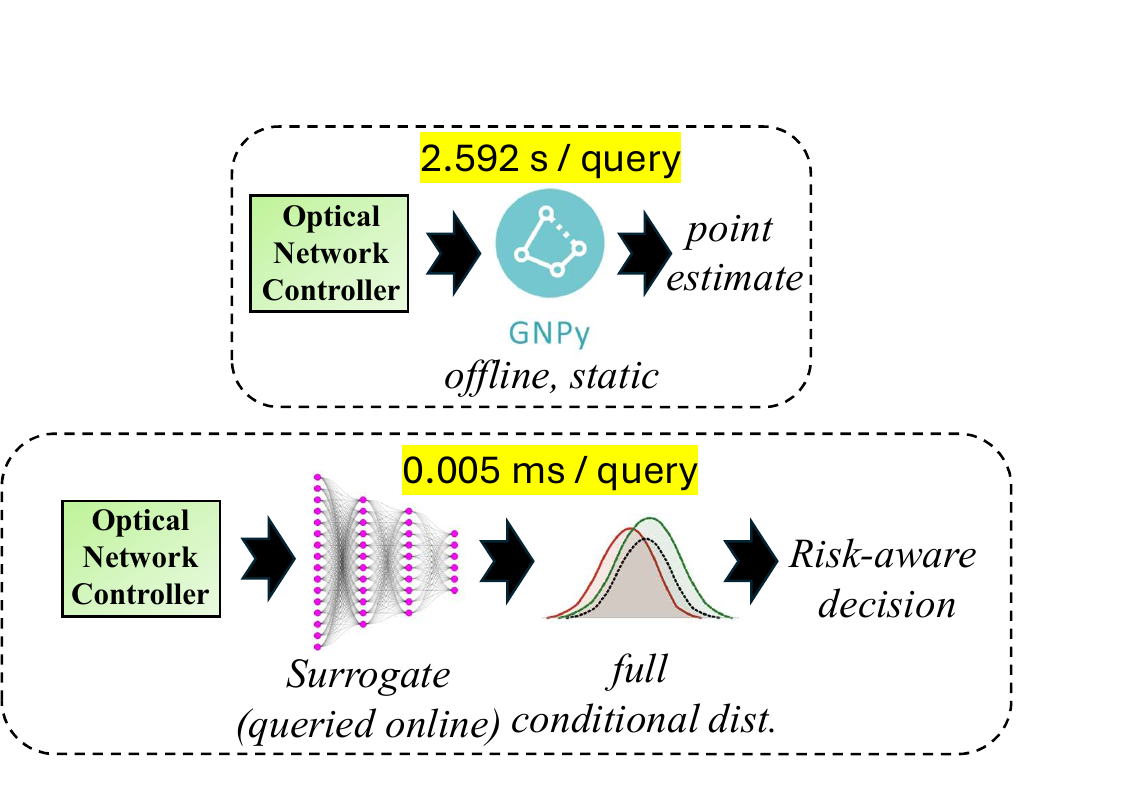}
    \caption{Existing approaches (top-half) query an offline simulator such as the GNPy at the provisioning time. The simulator provides a point estimate following a multi-second delay (\texttt{2.592 sec} as we measured). Our proposed surrogate replaces this setup with a probabilistic queryable model that provides the controller with a full conditional distribution over physical-layer outcomes -- within a \texttt{sub-millisecond delay}. Our proposed online setup allows the controller to make risk-aware decisions at runtime (bottom-half).}
    \label{fig:intro}
\end{figure}

The main contributions of this paper are listed below. (1) First, we propose \emph{a generative surrogate for the optical layer in the form of a conditional generative model trained on simulator-derived traces.} This surrogate exposes the underlying physical-layer behavior as a fast, probabilistic query engine to the control plane -- thus allowing online decision-making. (2) Second, we present \emph{a simulator-only training pipeline that does not require any live optical hardware traces.} This makes the surrogate practical for early-stage evaluation and supports reproducibility. (3) Third, we present \emph{the first case study in fast path feasibility estimation using the proposed surrogate.} We compare the surrogate against simulators, deterministic ML predictors, and quantile-regression baselines across nine research questions -- including integration into a lightweight RSA-lite controller with first-fit spectrum assignment. (4) We show that the generative risk-aware path-selection policy achieves the lowest blocking probability (BP) and the highest spectrum utilization among all four evaluated policies. We also show that \emph{the surrogate exposes a full conditional distribution over the physical-layer outcomes to the control plane, which quantile regressors cannot provide at runtime without retraining.} (5) We evaluate the proposed surrogate in a cross-emulator generalization setting against Mininet-Optical traces and show that feasibility rank ordering partially survives the domain shift with AUC~0.667 -- even when the Q-margin dynamic range differs by 2.61$\times$ (see \textbf{RQ5} in Sec.~\ref{sec:results}). (6) We demonstrate proactive dynamic reconfiguration under amplifier aging (\textbf{RQ6}) to show feasibility under gradual physical-layer degradation inherent to long-lived optical connections in production networks~\cite{vela2017qot,mahajan2021retraining}. We show that \emph{the Mixture Density Network (MDN)-based controller eliminates connection failures by triggering rerouting 19 steps earlier than the reactive baseline. The risk budget $\tau$ is a single tunable parameter for trading safety against unnecessary reroutes -- this is a capability the deterministic baseline structurally cannot provide.} (7) We show that surrogate prediction accuracy is stable across four topologies from synthetic benchmarks to real 37-node backbones, with MAE staying flat at 0.72--0.77\,dB (\textbf{RQ7}). (8) We validate generalization from analytical Gaussian Noise (GN)-model training data to GNPy-simulated traces on NSFNet and Cost266 topologies. We show that MAE degrades by less than 0.10\,dB and AUC remains at 0.997--0.999 under both out-of-distribution and cross-topology transfer conditions (\textbf{RQ8}). (9) Finally, using 11,209 telemetry snapshots from a live production optical network~\cite{alibaba2024} to derive the input feature vectors, we show a sub-millisecond latency advantage which is $6{,}275\times$ speedup over the simulator-based approach (GNPy), which routinely incurs multi-second latency (\textbf{RQ9}).
\begin{table}[t]
\centering
\caption{Comparison of optical-layer estimation approaches.}
\label{tab:approach-comparison}
\renewcommand{\arraystretch}{1.1}
\setlength{\tabcolsep}{3pt}
\small
\begin{tabular}{lccc}
\toprule
\textbf{Approach} & \textbf{Output} & \textbf{Full Dist.?} & \textbf{Snap.-Dep.?} \\
\midrule
GNPy / simulator      & Point est.      & \xmark  & Yes (explicit)  \\
Analytical GN/EGN     & Point est.      & \xmark  & Yes (calib.)    \\
Deterministic ML      & Point est.      & \xmark  & Yes (train.)    \\
Quantile regression   & Fixed quantiles & Partial & Yes (train.)    \\
\textbf{Ours}         & \textbf{Full cond.\ dist.} & \cmark & \textbf{No} \\
\bottomrule
\end{tabular}
\end{table}

\paragraph{\textbf{Reproducibility}} All experiments presented in this paper were conducted using the GNPy~\cite{gnpy} as the optical-layer simulator. The training data, code for the surrogate (VAE and MDN), the baselines, and testing scripts are available at \url{https://github.com/saisumanv/CoNEXT26-gen-surr}.

\section{Preliminaries}\label{sec:prelims}

\subsection{Optical control requires physical-layer reasoning}
Optical network controllers cannot make decisions based on the underlying network topology alone. For example, a network path that may be short in hop length may still offer poor performance if it passes through unfavorable spans~\cite{azodolmolky2009survey}, experiences high nonlinear interference~\cite{poggiolini2012gn,zhao2015nonlinear}, lacks sufficient OSNR~\cite{gnpy}, or has too little physical-layer margin to tolerate modeling uncertainty and evolving network conditions~\cite{sartzetakis2019accurate,mahajan2021retraining}.

\subsection{Existing abstractions are either slow or too rigid}
The current optical control stack~\cite{rafique2018mlautomation} is too rigid or static and finds it difficult to match their network control decision-making with the fast and highly dynamic traffic, changing spectrum occupancy, and transient physical- and network-layer conditions~\cite{sartzetakis2019accurate,pointurier2021mlqot,dycrono}. One option is to run detailed physical-layer simulations during the planning stage~\cite{gnpy}. Although this provides useful fidelity, it matches poorly to dynamic control, where several candidate network paths and configurations need to be evaluated quickly online~\cite{rafique2018mlautomation}. Another option is to adopt analytical models or even conservative physical layer margins~\cite{sartzetakis2019accurate,pointurier2021mlqot} -- although these are faster than running full detailed optical layer simulations, they only provide a point estimation, which can lead to overly conservative decisions and hence may miss provisioning opportunities, or may rely on risk-blind routes~\cite{sartzetakis2019accurate,pointurier2021mlqot}.



\section{Related Work}\label{sec:related}

\paragraph{Physical-layer estimation and QoT prediction.} Estimating the Quality of Transmission (QoT) of optical lightpaths has been studied using both analytical models and learned predictors. The Gaussian Noise (GN) model~\cite{poggiolini2012gn} provides closed-form OSNR estimates under uncompensated coherent transmission and remains the foundation for physics-based control-plane decisions. GNPy~\cite{gnpy} operationalizes the GN model as an open-source simulation platform used in current optical standards. The authors in~\cite{sartzetakis2019accurate} show that deterministic ML predictors trained on simulator traces can match or exceed analytical model accuracy on point-estimate QoT tasks. The authors in~\cite{pointurier2021mlqot} survey ML techniques for QoT estimation across regression, classification, and transfer learning settings. The authors in~\cite{rottondi2018ml} propose an ML method for QoT prediction of unestablished lightpaths using features extracted from neighboring established connections. The authors in~\cite{morais2018ml} study ML models for OSNR estimation in DWDM networks and report results across multiple topology sizes. The authors in~\cite{mata2018ai} survey AI methods for optical transmission and networking, covering regression, classification, and reinforcement learning approaches. The authors in~\cite{vela2017qot} study physical-layer degradation modeling for dynamic optical networks and show that learned models can track degradation trends that analytical models miss. We categorize all of these works as deterministic QoT predictors. Deterministic QoT predictors output a point estimate of the physical-layer metric and do not provide a distribution over possible outcomes. This forces the control plane to rely on conservative fixed margins to account for uncertainty that is never explicitly modeled. Our surrogate directly addresses this limitation by exposing a full conditional distribution over physical-layer outcomes rather than a point estimate.

\paragraph{Uncertainty quantification for optical networks.} The authors in~\cite{mahajan2021retraining} study retraining strategies for QoT estimators under distribution shift and show that static models degrade as network conditions evolve. The authors in~\cite{delezoide2019probabilistic} propose probabilistic QoT models that produce confidence intervals over margin estimates. The authors in~\cite{panayiotou2019uncertainty} study margin uncertainty in elastic optical networks and show that deterministic margin allocation leads to spectrum over-provisioning. Quantile regression~\cite{koenker1978regression,koenker2001quantile} has been applied to QoT estimation as a computationally lightweight alternative that provides fixed confidence intervals at training-specified quantiles. Conformal prediction~\cite{angelopoulos2023} provides distribution-free coverage guarantees and has been applied to network traffic classification~\cite{comnetSI26}. We categorize all of these works as partial uncertainty methods. Partial uncertainty methods provide fixed summaries of the distribution -- selected quantiles or confidence intervals -- but the target quantiles or coverage levels must be specified at training time. This means the controller can only evaluate candidate paths under a fixed risk budget that was decided offline. Our surrogate is different from these works because the MDN produces a full sampleable conditional distribution that the controller can query at arbitrary violation thresholds at runtime, without retraining for each decision.

\paragraph{Surrogate and emulator models.} Surrogate modeling -- replacing expensive simulators with fast learned approximations -- is well established in computational science and engineering~\cite{forrester2008surrogates}. The authors in~\cite{raissi2019pinn} propose physics-informed neural networks that embed physical constraints into the learning objective to improve generalization under distribution shift. The authors in~\cite{rusek2019routenet} apply a graph neural network-based surrogate to packet-level network simulation, showing that learned surrogates can replace discrete-event simulators for network performance prediction. These works establish the surrogate modeling paradigm in scientific computing and networking, but none of them has been applied to optical physical-layer simulation. Our work is different from these works because we are, to the best of our knowledge, the first to frame the optical-layer simulator as a surrogate modeling problem, and the first to expose the resulting model as a probabilistic query engine to the control plane.

\paragraph{Digital twins and online control.} The authors in~\cite{xie2021digitaltwin} propose a digital twin framework for optical networks that maintains a continuously updated virtual replica of the physical layer to support control-plane decisions. The authors in~\cite{rafique2018mlautomation} survey machine learning for network automation and identify fast, uncertainty-aware physical-layer interfaces as a key open problem. Our surrogate can be viewed as a lightweight instantiation of the digital twin concept -- a probabilistic model trained offline on simulator traces and queried online by the controller -- with the offline--online separation making it practical for early-stage deployment without live hardware instrumentation. This is different from existing digital twin works, which require continuous measurement feedback and real-time model updates.

\paragraph{Routing and spectrum assignment.} The authors in~\cite{chatterjee2015rsa} survey routing and spectrum assignment in elastic optical networks across heuristic, exact, and learning-based methods. The authors in~\cite{gerstel2012elastic} introduce elastic optical networking and motivate the need for fine-grained spectrum management. Physical-layer-aware routing algorithms incorporate impairment constraints into path selection~\cite{azodolmolky2009survey, zhao2015nonlinear}, but rely on deterministic feasibility checks or conservative margin rules that do not expose uncertainty to the controller. Our generative risk-aware policy is different from these works because it integrates violation probability estimates directly into path ranking, which improves spectrum utilization and reduces fragmentation relative to deterministic baselines without increasing blocking probability.

\paragraph{Dynamic reconfiguration under degradation.} The authors in~\cite{vela2017qot} show that span degradation and amplifier aging are recognized challenges for long-lived optical connections and that soft failure localization requires early detection before performance drops below threshold. The authors in~\cite{mahajan2021retraining} show that QoT estimator accuracy degrades over time as physical conditions evolve, motivating the need for proactive rather than reactive responses. Existing works on proactive reconfiguration rely on threshold crossing of point estimates to trigger rerouting~\cite{sartzetakis2019accurate, pointurier2021mlqot}. This is limited because a point-estimate predictor cannot distinguish between a path that is confidently above threshold and one that is marginally above threshold with high uncertainty -- both receive the same reconfiguration decision. Our MDN-based proactive controller addresses this limitation directly. By monitoring violation probability rather than predicted Q-margin, it triggers rerouting an average of 19 steps earlier than the reactive baseline and eliminates connection failures, with the risk budget $\tau$ as a single tunable parameter for trading safety against unnecessary reroutes.

\section{Generative Surrogate for the Optical Layer}\label{sec:design}
\begin{figure*}[htp]
\centering
    \includegraphics[width=0.75\linewidth,trim={0.075cm 5.75cm 4.0cm 1.25cm},clip]{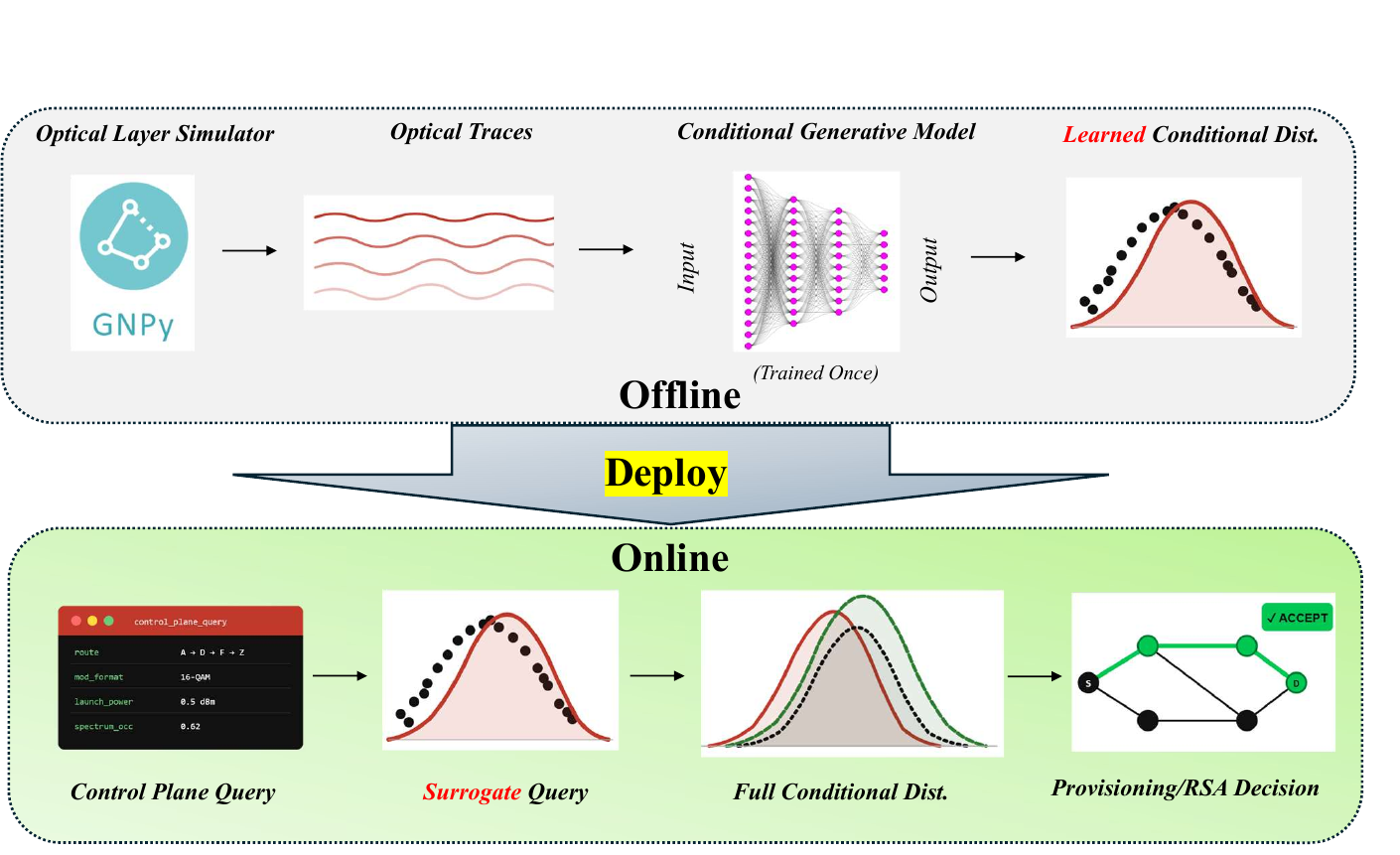}
    \caption{Overview of the generative surrogate framework. \textit{Offline}: traces from a validated optical simulator (GNPy) are used to train a conditional generative model once. \textit{Online}: the trained surrogate is queried by the control plane to produce a full conditional distribution over physical-layer metrics, which eventually drives provisioning and RSA decisions.}
    \label{fig:surrogate}
\end{figure*}

Figure~\ref{fig:surrogate} shows the design of our proposed generative probabilistic surrogate for the optical layer. The surrogate is first trained offline (top rectangle) and queried online i.e, post-deployment (bottom rectangle). The online--offline separation is important as it exports the expensive simulations which cannot be realistically done at runtime. The simulations remain useful in our framework, but they are moved out of the critical path of control-plane decision-making (routing/spectrum assignment/provisioning/reconfiguration).

\subsection{Inputs, Outputs, and Candidate Models}

The surrogate is conditioned on features that describe the candidate optical state -- the features are listed in the first-row in Tab.~\ref{tab:surrogate_design_space}. The exact input feature set may vary depending on the simulator support and the target control plane tasks. The surrogate outputs a conditional distribution over physical-layer quantities that are found to be relevant to optical-layer control as shown in the second-row of Tab.~\ref{tab:surrogate_design_space}. The third-row presents the candidate generative models that were considered in this work. 
\begin{table*}[t]
\centering
\caption{Design space for a generative surrogate of the optical layer.}
\label{tab:surrogate_design_space}
\small
\begin{tabular}{p{0.20\textwidth} p{0.72\textwidth}}
\hline
\textbf{Component} & \textbf{Examples} \\
\hline
\textbf{Input features} &
Path length and span count; per-span distance or aggregate distance
statistics; modulation format; baud rate or symbol rate; launch power;
channel spacing; spectrum occupancy around the candidate channel;
traffic load or utilization; amplifier noise parameters;
topology-level features; candidate route identifier or path embedding. \\
\hline
\textbf{Output quantities} &
OSNR; Q-factor or Q-margin; nonlinear interference penalty; feasibility
probability; outage or threshold-violation probability; confidence
interval over margin. \\
\hline
\textbf{Candidate models} &
Conditional variational autoencoders; mixture density networks;
normalizing flows; diffusion models. \\
\hline
\end{tabular}
\end{table*}
For a first case study, we chose a mixture density network as the primary surrogate. A conditional variational autoencoder (VAE) was also evaluated as an alternative; results are reported in Appendix~\ref{app:vae}.  Here, the candidate model choice is far less important than the systems-level abstraction: \emph{the optical layer is represented as a fast conditional distribution that can be queried by the control plane at runtime.}

\subsection{Why not only quantile regression?}
Quantile regression is a well-known practical and computationally lightweight alternative to run a full generative model~\cite{koenker1978regression,koenker2001quantile}. However, quantile regression and its variants only provide a limited set of conditional quantiles rather than a model that can be sampled for fetching the target optical-layer outcome distribution~\cite{koenker2001quantile,bishop1994mdn}. Following this, we categorize three cases where quantile regression falls behind in realizing a fully online interface between the physical-layer and the optical control plane. (i) First, since Quality of Transmission (QoT) is evaluated before provisioning or upgrading optical connections, the path feasibility decisions often depend on threshold-crossing probabilities~\cite{sartzetakis2019accurate,pointurier2021mlqot}. On the other hand, a generative surrogate can estimate these probabilities directly by sampling from the  learned conditional distribution~\cite{sohn2015conditional,bishop1994mdn} provided by the learned model. (ii) Second, physical-layer uncertainty provided by quantile regressors may be asymmetric, heavy-tailed, or even multimodal across operating regimes. While it has been shown that a small number of quantiles may summarize this uncertainty, they may still not preserve its shape~\cite{koenker2001quantile,bishop1994mdn,rezende2015normalizing}. The tail behavior and the shape of the distribution affects whether a candidate lightpath is admitted, rejected, or assigned an additional margin. So, even if the expected safety margin -- the safety gap between the predicted physical-layer signal quality of a lightpath and the minimum signal quality required for reliable transmission -- looks acceptable in reality, the controller may reject that path if the learned distribution shows a practically meaningful probability of very low OSNR/Q or threshold violation. (iii) Third, control-plane decisions need to be made by comparing many candidate paths under different risk budgets or tolerances. Although quantile regressors can report selected confidence levels, they do not provide a reusable sampled distribution that can be assessed under different risk tolerances. This specific limitation applies even when the quantile regressor is trained on a small fixed set of quantiles (e.g., P10, P50, P90). This is the same setting we evaluate in this paper. However, a quantile regressor that is trained on a dense grid of quantiles, or one that interpolates between the trained quantiles, can potentially narrow this gap, but only at the cost of additional training and inference.

\begin{figure}[t]
\centering
\includegraphics[width=0.65\columnwidth]{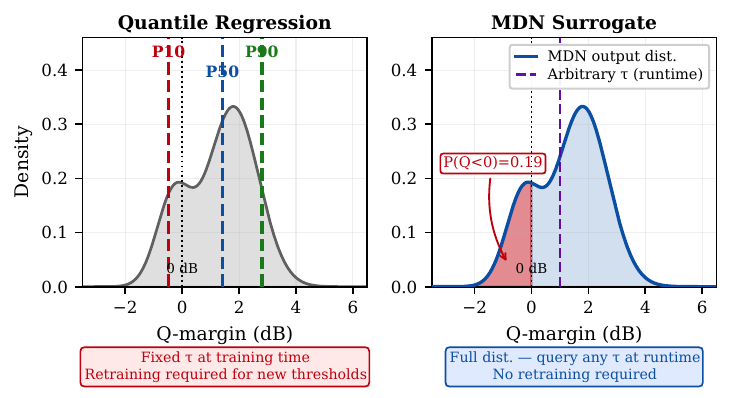}
\caption{Quantile regression exposes only fixed pre-specified quantiles (left), requiring retraining for each new risk threshold. The MDN surrogate exposes a full sampleable conditional distribution (right), allowing the controller to query violation probability at arbitrary thresholds $\tau$ at runtime without retraining.}
\label{fig:qr_vs_mdn}
\end{figure}

Our claim here is not that quantile regression is weak for control plane decision-making. We believe it is a strong and practical alternative~\cite{koenker1978regression,koenker2001quantile}. Our claim, instead, is that a sampleable surrogate gives the controller a reusable interface for expected margin, violation probability, tail risk, and risk-aware ranking~\cite{sartzetakis2019accurate,pointurier2021mlqot,lakshminarayanan2017deepensembles}.

\section{Simulator-Only Training Pipeline}\label{sec:pipeline}
Our proposed pipeline (Fig.~\ref{fig:surrogate}) does not require any live optical network hardware during control plane decision-making. Instead, it relies on simulator-generated traces, which makes it practical for early-stage evaluation and reproducible experimentation.

\subsection{Trace generation}
To generate the traces, we apply two steps. First, we create a set of candidate optical states by sampling topologies, paths, traffic loads, spectrum assignments, modulation formats, launch-power settings, and span-level conditions. Second, each sampled optical state is evaluated using an optical-layer simulator, namely, GNPy~\cite{gnpy}, which is used to produce the target physical-layer outcomes. A training tuple has the form $(x, y)$, where \(x\) is the candidate optical/network state and \(y\) is the physical-layer outcome derived from the simulator. In our setting, \(x\) may include path, topology, modulation, launch power, and spectrum-occupancy features (first-row in Tab.~\ref{tab:surrogate_design_space}), while \(y\) may include OSNR, Q-margin, nonlinear penalty, and feasibility (second-row in Tab.~\ref{tab:surrogate_design_space}).

\subsection{Dataset Construction}
We construct the dataset to cover both feasible and infeasible paths, low-margin and high-margin paths, and multiple operating regimes, including short-, medium-, and long-haul paths; low-, moderate-, and high-load traffic conditions; sparse and dense spectrum occupancy; low, nominal, and high launch-power settings; different modulation formats and baud rates; linear-noise-dominated and nonlinear-interference-dominated regimes; favorable and unfavorable span configurations; nominal and degraded amplifier/noise conditions; and both low-margin and high-margin feasibility regions~\cite{gerstel2012elastic,chatterjee2015rsa,poggiolini2012gn,gnpy,pointurier2021mlqot}. The surrogate is only useful if it can generalize across a range of states that the controller may query at runtime. For this reason, we diversify the dataset via varying route length, span count, launch power, spectrum occupancy, modulation format, offered load, and amplifier or noise assumptions~\cite{chatterjee2015rsa,poggiolini2012gn,gnpy,pointurier2021mlqot}. In our study, we chose the sampling ranges to account for both sides of the feasibility boundary instead of matching the query distribution of any specific operational network. This aligns with the philosophy that in practice operators who deploy the surrogate will need to construct the training population from the parameter ranges of their own network. This is what we do for the Alibaba network in RQ9, our chosen production candidate.

\subsection{Training Objective}
For a conditional generative model, the training objective is to learn the conditional distribution: $p(y \mid x),$ where \(x\) is the candidate optical state and \(y\) is the physical-layer outcome. For a VAE, this can be learned using a reconstruction term and latent regularization. For a mixture density network, the model can directly learn a mixture distribution over the output variables. For a quantile or distributional model, the objective can encourage calibrated prediction intervals.

\section{Results and Discussion}\label{sec:results}
Given a candidate lightpath, the controller needs to know if that candidate path satisfies the physical-layer threshold by querying our generative surrogate, which then returns a distribution over path quality. This allows the controller to choose one of the several decision rules:
\begin{itemize}
\item accept a path if its expected margin exceeds a threshold;
\item accept a path only if the probability of violation is below a risk budget;
\item prefer the candidate path with the lowest predicted violation probability;
\item trade physical-layer risk against spectrum fragmentation or routing cost.
\end{itemize}

We evaluated our generative surrogate along nine research questions, RQ1--RQ9, defined below. We compared the surrogate against three baselines, namely, (i)~simulator outputs, (ii)~deterministic ML predictors, and (iii)~quantile-regression models. For continuous outputs such as OSNR and Q-margin, we measured mean absolute error (MAE) and root mean squared error (RMSE). For feasibility classification, we measured accuracy, precision, and recall.

Additionally, we measured whether the surrogate is useful in an online setting as an optical-layer query engine. We measured single query latency, time taken to evaluate a candidate path set, and violation-probability calibration near the feasibility threshold. We also measured whether the violation probability estimate changes which path the controller selects. For RQ5--RQ9, we measured cross-emulator generalization, proactive rerouting under span degradation, topology-scale stability, and GNPy-validated transfer across backbone topologies, and latency of inference on production network feature vectors.

\subsection{RQ1 — Accuracy of Physical‑Layer Estimation: Surrogate approximates simulator outputs}
We first compared the surrogate against the simulator outputs to test whether the surrogate can estimate  OSNR, Q-margin, nonlinear penalty, and the final control-plane decision with sufficient accuracy. We then compare it against deterministic ML predictors and quantile regressors.
\begin{table}[!ht]
\centering
\caption{Accuracy of physical-layer estimation against simulator-derived labels.}
\label{tab:rq1_accuracy}
\small
\renewcommand{\arraystretch}{1.08}
\begin{tabular}{lcccc}
\toprule
\textbf{Model} & \textbf{OSNR} & \textbf{Q-margin} & \textbf{Feas.} & \textbf{Feas. Recall} \\
 & \textbf{MAE} & \textbf{MAE} & \textbf{Acc.} &  \\
\midrule
Deterministic ML & 0.748 & 0.749 & 0.951 & 0.950 \\
Quantile regression & -- & 0.785 & -- & -- \\
Generative surrogate & -- & 0.787 & 0.948 & 0.944 \\
\bottomrule
\end{tabular}
\end{table}
We note that this is an acceptable tradeoff because the surrogate exposes a conditional distribution, which allows the controller to estimate the threshold-violation probability instead of solely relying on point prediction provided by deterministic MLP.

\subsection{RQ2 — Latency vs. Simulator Calls: Surrogate is faster}\label{subsec:rq2}
Next, we compared the surrogate over the latency of each call against the simulator. Here, the metric is not just latency per call, but also the time taken to evaluate the candidate path set -- because in reality, the controller rarely evaluates a single candidate path.
\begin{table}[t]
\centering
\caption{Candidate-set evaluation time for simulator calls and learned surrogate inference.}
\label{tab:rq2_latency}
\renewcommand{\arraystretch}{1.08}
\setlength{\tabcolsep}{3pt}
\begin{tabular}{lcccc}
\toprule
\textbf{Method} & \textbf{1} & \textbf{10} & \textbf{100} & \textbf{1000} \\
 & \textbf{path} & \textbf{paths} & \textbf{paths} & \textbf{paths} \\
\midrule
GNPy simulator      & 2592 ms & 25.9 s  & 259 s   & 43.2 min \\
Deterministic ML    & 0.007 ms & 0.07 ms & 0.72 ms & 7.21 ms \\
Quantile reg.       & 0.011 ms & 0.11 ms & 1.10 ms & 11.0 ms \\
Generative surr.    & 0.005 ms & 0.05 ms & 0.53 ms & 5.34 ms \\
\bottomrule
\end{tabular}
\end{table}
We measured the runtime for a transmission example using GNPy simulator, which was found to be \texttt{2.592 sec} for a single path query, as shown in Tab.~\ref{tab:rq2_latency}. We must note that this is distinct from the physical propagation latency that GNPy reported for the optical path.

\subsection{RQ3 — Uncertainty Quality vs. Quantile Regression: Calibrated MDN matches quantile coverage while exposing violation risk}
We tested whether the surrogate is able to provide more useful uncertainty information compared to quantile regressors. This evaluation lets us know if a full conditional distribution is more useful than a fixed set of conditional quantiles.

\begin{table}[t]
\centering
\caption{Uncertainty quality for Q-margin estimation.}
\label{tab:rq3_uncertainty}
\small
\setlength{\tabcolsep}{3pt}
\renewcommand{\arraystretch}{1.08}
\begin{tabular}{lccc}
\toprule
\textbf{Model} & \textbf{Cov.} & \textbf{Calib. err.} & \textbf{Width} \\
\midrule
Quantile reg. & 0.781 & 0.020 & 2.522 dB \\
MDN + cal.    & 0.770 & 0.031 & 2.498 dB \\
\bottomrule
\end{tabular}
\end{table}
We found that the calibrated MDN surrogate showed P10--P90 interval coverage comparable to the quantile-regression baseline, with similar interval width and calibration error, as shown in Tab.~\ref{tab:rq3_uncertainty}. Quantile regression showed slightly higher coverage -- however, this comparison is made on pre-specified quantiles where quantile regression is specifically optimized to perform well. This retraining limitation was not found to apply to the MDN surrogate. The MDN exposes a full conditional distribution that can be queried at arbitrary thresholds at runtime without retraining.


\begin{table*}[t]
\centering
\caption{RSA-lite controller-level path-selection results with first-fit spectrum assignment.}
\label{tab:rq4_controller}
\renewcommand{\arraystretch}{1.08}
\setlength{\tabcolsep}{4pt}
\small
\begin{tabular}{lcccccc}
\toprule
\textbf{Policy} & \textbf{Block.} & \textbf{Viol.} & \textbf{Margin} & \textbf{Util.} & \textbf{Frag.} & \textbf{Time} \\
\midrule
Shortest feasible + FF       & 0.405 & 0.000 & 3.860 dB & 0.643 & 0.112 & 5.844 s \\
Det.\ best-margin + FF        & 0.397 & 0.054 & 4.262 dB & 0.646 & 0.113 & 0.016 ms \\
Conservative margin + FF     & 0.454 & 0.021 & 4.269 dB & 0.570 & 0.066 & 0.016 ms \\
Generative risk-aware + FF   & 0.395 & 0.054 & 4.144 dB & 0.652 & 0.100 & 0.012 ms \\
\bottomrule
\end{tabular}
\end{table*}

\subsection{RQ4 — Impact on Controller‑Level Decisions: Simple controller-level decision}
Finally, we evaluated whether the surrogate improves the control-plane decision-making, not just the physical-layer behavior. To test this, we integrated the surrogate into a simple baseline routing and path-selection algorithm. When the controller receives a request with a set of candidate paths  \(P_{s,d}\), it selects one path based on the chosen path-selection policy-- the surrogate first estimates the probability \(\rho(p)\) that each path in the set \(P_{s,d}\) falls below the required OSNR/Q level for the chosen modulation:
\[
\rho(p) = \Pr(Q(p) < Q_{\min}(m)).
\]
Following this, the controller selects the candidate path with the smallest violation probability given by:
\[
p^* = \arg\min_{p \in P_{s,d}} \rho(p),
\]
subject to spectrum availability.

We compared this path-selection policy against three baselines, namely, shortest feasible path, deterministic best-margin path, and conservative-margin path selection (see first-column in Tab.~\ref{tab:rq4_controller}).

We integrated the surrogate into a lightweight controller that does path selection, as a proof-of-concept. For every new connection request, this controller receives a set of candidate paths between the source and destination. The controller then queries the surrogate -- the learned optical-layer model -- for each path in the candidate set. Upon query, the surrogate returns a predicted Q-margin distribution per candidate path, from which the controller can estimate the probability that specific path violates the required OSNR/Q threshold.

This violation probability is used to rank all the paths in the candidate set before making path provisioning decision. As shown in Tab.~\ref{tab:rq4_controller}, the generative risk-aware policy showed the highest spectrum utilization (0.652) among all the policies. Additionally, the spectrum fragmentation was found to be lower (0.100) compared to the deterministic best-margin (0.113) and shortest feasible (0.112) baselines. In this risk-aware policy, paths with lower violation or violation probability are given priority, while paths with violation probability above the threshold are rejected. This setup allows the controller to make fast provisioning decisions at runtime using a probabilistic physical-layer signal instead of labeling each candidate path with a single feasibility label or relying on a deterministic offline/static simulator call for each candidate path. We must note that we found the improvements in blocking probability and spectrum utilization over the deterministic best-margin policy to be small (0.395 vs.\ 0.397 and 0.652 vs.\ 0.646, respectively). These results are from a single run. We will confirm these improvements in our future work by evaluating across multiple seeds with variance bounds.

After the controller selects a path, it assigns spectrum using first-fit contiguous slot allocation over every link on that chosen path. At this stage, the controller blocks the request if no candidate path satisfies the policy constraint or if the chosen path cannot support a contiguous slot block.

\subsection{RQ5 — Cross-emulator generalization (Mininet-Optical)}\label{subsec:mininet}

The pipeline described in Sec.~\ref{sec:pipeline} trains and evaluates exclusively on GNPy-generated traces. To test whether the same learned surrogate generalizes beyond its training emulator, we evaluated it against 2{,}000 traces produced by the  Mininet-Optical~\cite{mininetoptical} -- a well-known network-level emulator that models ROADM switching, amplifier cascades, and spectrum assignment independent of GNPy's underlying analytical propagation model.

\paragraph{Setup.} Mininet-Optical exposes  a physical-layer feature set ($n_\text{spans}$, span length, launch power, noise figure). However, the training of the surrogate involved the full topology-aware feature vector. To address this discrepancy to an extent, we retrained a reduced surrogate on the three features computable from both emulators: number of spans, path length (derived as $n_\text{spans} \times \ell_\text{span}$), and launch power. Q-margin labels from Mininet-Optical are derived as $\text{gOSNR}_\text{dB} - \theta_\text{rx}$, where $\theta_\text{rx} = 20$\,dB is the fixed receiver threshold.

\paragraph{Results.} Tab.~\ref{tab:mininet} presents the MAE, RMSE, feasibility accuracy, AUC, and $\mathrm{P}_{10}$--$\mathrm{P}_{90}$ interval coverage on the held-out GNPy test set and the Mininet-Optical traces. We found that the MDN surrogate achieved AUC~0.889 on GNPy and AUC~0.667 on Mininet. This indicates that feasibility rank ordering partially survived the domain shift between the emulators. Additionally, we found that the absolute Q-margin MAE degraded from 2.74\,dB to 11.31\,dB. This degradation is driven primarily by a distributional width mismatch -- the GNPy training distribution has a standard deviation of 5.47\,dB while the Mininet-Optical had 14.31\,dB (a 2.61$\times$ ratio), rather than a systematic mean shift (1.20\,dB).

\begin{figure}[t]
\centering
\includegraphics[width=0.65\columnwidth]{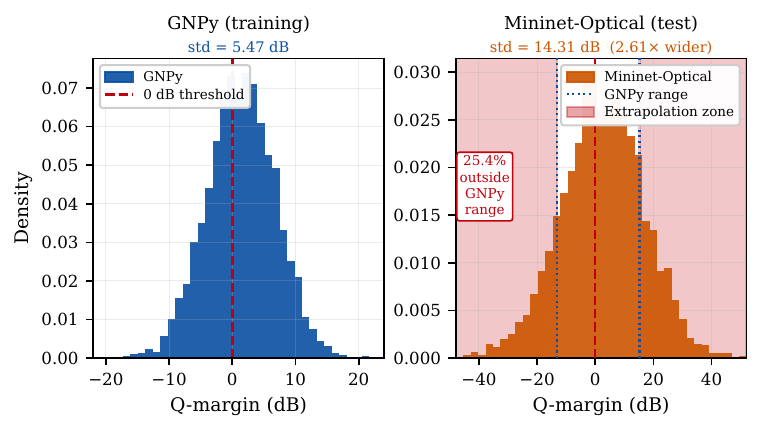}
\caption{Q-margin distributions for GNPy (training, left) and Mininet-Optical (test, right). The Mininet distribution is 2.61$\times$ wider (std 14.31\,dB vs.\ 5.47\,dB), and 25.4\% of Mininet traces fall outside the GNPy training range entirely (shaded), placing them in a hard extrapolation regime.}
\label{fig:rq5_dist}
\end{figure}

We must note that 25.4\% of Mininet traces fall entirely outside the Q-margin range observed during this training, which places them in a hard extrapolation regime. Finally, the interval coverage collapsed from 0.84 to 0.25, which confirms that the uncertainty calibration does not transfer across emulators without adaptation. This means that the controller should not trust the absolute violation probabilities for admission decisions under such a collapse of the coverage. Having said that, the surrogate can still be used to rank the candidate paths, since the rank ordering itself was found to partially survive (AUC~0.667).

\paragraph{Implications.} These results identify two concrete gaps for future work. First, training set augmentation with span configurations covering a wider Q-margin dynamic range is required. Second, lightweight post-hoc recalibration (e.g.,~temperature scaling or conformal prediction~\cite{angelopoulos2023}) is needed to restore the coverage on out-of-distribution traces like we observed when training the models under Mininet-Optical. Interestingly, the partial preservation of rank ordering indicated by AUC~0.667 suggests that the surrogate's latent representation captures physically meaningful structure even under emulator shift. This provides a foundation for domain-adaptive fine-tuning with few Mininet samples.
\begin{table}[t]
\centering
\caption{Cross-emulator generalization: GNPy (held-out) vs.\ Mininet-Optical. Reduced feature set: $\{n_\text{spans},\, \ell_\text{path},\, P_\text{launch}\}$.}
\label{tab:mininet}
\small
\setlength{\tabcolsep}{3pt}
\begin{tabular}{lrrrrcc}
\toprule
Model & MAE & RMSE & Acc. & AUC
      & $\mathrm{Cov}_{10\text{--}90}$ & Width \\
\midrule
MLP (GNPy)     & 2.74  & 3.24  & 0.802 & 0.802 & --    & --       \\
MLP (Mininet)  & 11.68 & 14.52 & 0.492 & 0.527 & --    & --       \\
\midrule
MDN (GNPy)     & 2.74  & 3.24  & 0.799 & 0.889 & 0.842 & 9.32\,dB \\
MDN (Mininet)  & 11.31 & 14.33 & 0.516 & 0.667 & 0.246 & 8.64\,dB \\
\bottomrule
\end{tabular}
\end{table}

\subsection{RQ6 — Dynamic reconfiguration under span degradation}
\label{subsec:dynreconfig}
Optical control planes must respond to gradual physical-layer degradation~\cite{sartzetakis2019accurate,pointurier2021mlqot}. For example, amplifier aging often raises the effective noise figure (NF) of a span over time. To this end, we evaluated whether the generative surrogate's violation probability estimate allows for \emph{proactive} rerouting that can prevent connection failures, as opposed to a deterministic surrogate that only reacts once the predicted Q-margin crosses zero.

\paragraph{Setup.} We simulated the NF drift on $N{=}200$ lightpaths over $T{=}100$ discrete time steps with drift rate $\alpha{=}0.05$\,dB/step and additive jitter $\mathcal{N}(0,\,0.05^2)$\,dB, which gives a maximum degradation of 5\,dB. We initialized the lightpaths with ground-truth Q-margin in $[0.5,\,3.0]$\,dB to ensure that all paths degrade below the feasibility threshold within the simulation horizon. We trained the surrogate on the four observable control-plane features $\{n_\text{spans},\,\ell_\text{path},\,P_\text{launch},\,\text{NF}\}$ against full Q-margin labels that also depend on unobservable route and load parameters. This was deliberately designed to create genuine residual uncertainty (MAE\,=\,0.91\,dB) that the MDN captures as a predictive distribution but the deterministic MLP collapses to a point estimate.

We compared the two controllers:
\begin{itemize}
  \item \textbf{Reactive (MLP):} triggers rerouting when
        $\hat{q}_\text{margin} < 0$\,dB.
  \item \textbf{Proactive (MDN, $\tau$):} triggers rerouting when
        $P(\text{violation}) > \tau$, \emph{before} the predicted
        mean crosses zero.
\end{itemize}

A \emph{violation} is recorded when the true Q-margin drops below 0\,dB before rerouting is triggered. A \emph{false alarm} is recorded when rerouting is triggered but the true Q-margin at that moment exceeds 1\,dB (unnecessary reroute).

\begin{figure}[t]
\centering
\begin{minipage}[t]{0.45\columnwidth}
\centering
\includegraphics[width=\linewidth]{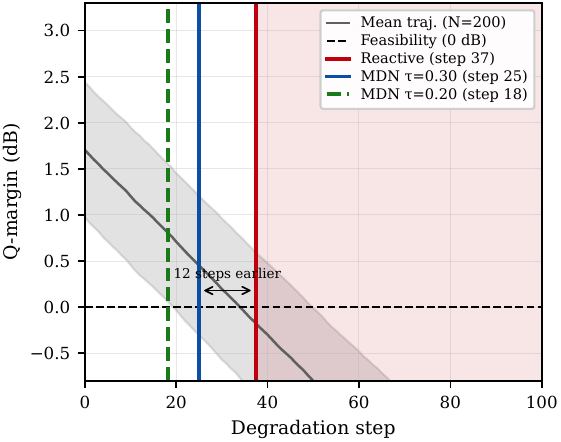}
\subcaption{Q-margin degradation trajectory.}
\label{fig:rq6_timeseries}
\end{minipage}
\hfill
\begin{minipage}[t]{0.515\columnwidth}
\centering
\includegraphics[width=\linewidth]{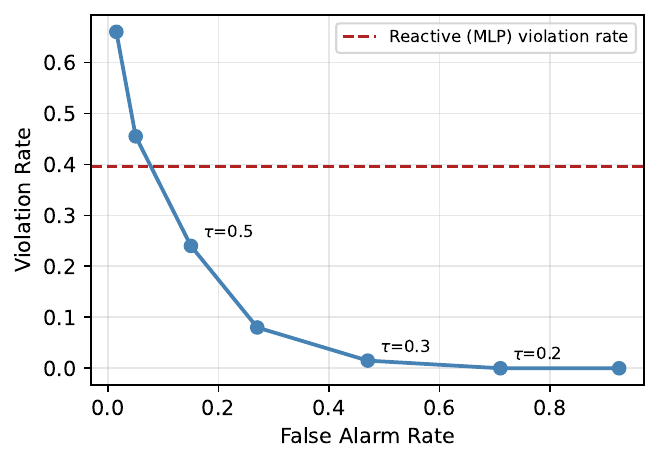}
\subcaption{Violation--false-alarm tradeoff across $\tau$.}
\label{fig:dynreconfig_pareto}
\end{minipage}
\caption{Dynamic reconfiguration under NF degradation ($N{=}200$ lightpaths, $\alpha{=}0.05$\,dB/step). Left: mean Q-margin trajectory with MDN proactive and reactive trigger points. Right: violation rate vs.\ false alarm rate across risk budgets $\tau \in [0.10, 0.70]$; the reactive MLP baseline (dashed) operates at a fixed 39.5\% violation rate with no tunable tradeoff.}
\label{fig:rq6_combined}
\end{figure}

\paragraph{Results.} Table~\ref{tab:dynreconfig} shows the results for the reactive baseline and two MDN operating points. Our results show that the reactive controller incurs a 39.5\% violation rate --- connections fail before the point estimate crosses the threshold --- and reroutes late at a mean Q-margin of only 0.51\,dB, thus leaving a minimal safety margin. On the other hand, at $\tau{=}0.20$, the MDN eliminated all violations but triggered rerouting 19.2 steps earlier on average, resulting in a 71\% false alarm rate. At $\tau{=}0.30$, violations dropped to 1.5\% while false alarms fell to 47\%, with rerouting occurring at a mean Q-margin of 1.03\,dB. This mean Q-margin provides a meaningful safety margin and avoids the excessive conservatism of lower thresholds. Figure~\ref{fig:dynreconfig_pareto} shows the full violation--false-alarm tradeoff across $\tau \in [0.10,\,0.70]$. The MDN's risk budget $\tau$ operates as a single tunable parameter that continuously interpolates between safety-first and efficiency-first operation. This is a capability the deterministic baseline structurally cannot provide. We must note that the reactive baseline triggers the rerouting only after the predicted Q-margin crosses 0\,dB, i.e., it reacts only at the feasibility threshold (rerouting decision is done late by design). This is in contrast with the deterministic baseline with a fixed safety offset (e.g., triggering at 1\,dB), which would reroute earlier than our reactive baseline. We leave the comparison of the MDN against such offset-based baselines, and also against a quantile-regression-based trigger for future work.

\begin{table}[t]
\centering
\caption{Dynamic reconfiguration under NF degradation. $N{=}200$ lightpaths, $T{=}100$ steps, $\alpha{=}0.05$\,dB/step.}
\label{tab:dynreconfig}
\small
\setlength{\tabcolsep}{3pt}
\begin{tabular}{lcccccc}
\toprule
Controller & Viol. & Viol.\ Rate & FA
           & FA Rate & Step & Q-mar. \\
\midrule
Reactive (MLP)         & 79/200 & 0.395 & 21/200  & 0.105 & 37.4 & 0.51\,dB \\
MDN $\tau{=}0.20$      & 0/200  & 0.000 & 142/200 & 0.710 & 18.2 & 1.38\,dB \\
MDN $\tau{=}0.30$      & 3/200  & 0.015 & 94/200  & 0.470 & 25.0 & 1.03\,dB \\
\bottomrule
\end{tabular}
\end{table}

\subsection{RQ7 — Surrogate Scalability Across Topologies}
\label{subsec:rq7}

We evaluated whether the surrogate's accuracy and uncertainty quality hold as topology size grows from synthetic benchmarks to real named backbones. We chose four topologies, namely, two synthetic regimes (topo-small, mean path 420\,km; topo-medium, mean path 850\,km) and two real backbone topologies --- NSFNet (14 nodes, 21 edges, US backbone) and Cost266 (37 nodes, 61 edges, pan-European backbone) --- using realistic edge lengths from published topology data~\cite{cost266,nsfnet}. We generated surrogate-compatible datasets for these four topologies. For each topology, we sampled 500 connection requests with 10 candidate paths each. We then trained independent MLP and MDN surrogates, and evaluated them on a held-out 20\% test set.

\paragraph{Results.} Table~\ref{tab:rq7_scale} summarizes MAE, feasibility accuracy, AUC, and P10--P90 interval coverage across all four topologies. MAE remained flat at 0.72--0.77\,dB across all topologies, which confirms that surrogate prediction accuracy does not degrade as topology size increases (from 14 to 37 nodes). AUC improved on the real topologies (0.998 on NSFNet, 0.997 on Cost266) relative to the synthetic benchmarks (0.980, 0.988). This is interesting because, it indicates a wider Q-margin dynamic range of long-haul real backbone paths which makes feasibility boundaries easier for the surrogates to learn. Interval coverage remained within 0.81--0.89 across all four topologies i.e., close to the nominal 0.80 target.

We note that NSFNet and Cost266 topologies exhibit very low feasibility rates (4\% and 3\% respectively) due to the long mean path lengths of real backbone topologies ($\sim$2900--3200\,km). However, the high AUC suggests that the surrogate is able to correctly identify the rare feasible paths despite this imbalance. Moreover, the feasibility accuracy is dominated by the majority infeasible class and thus it should be interpreted accordingly.

\begin{table}[t]
\centering
\caption{Surrogate scalability across topologies. MAE and AUC are for the MDN surrogate unless noted.}
\label{tab:rq7_scale}
\small
\setlength{\tabcolsep}{3pt}
\begin{tabular}{lrrrrrr}
\toprule
\textbf{Topology} & \textbf{N} &
\textbf{MAE\textsubscript{MLP}} &
\textbf{MAE\textsubscript{MDN}} &
\textbf{Acc} & \textbf{AUC} &
\textbf{Cov\textsubscript{10--90}} \\
\midrule
topo-small  & --  & 0.720 & 0.734 & 0.920 & 0.980 & 0.830 \\
topo-medium & --  & 0.752 & 0.758 & 0.954 & 0.988 & 0.815 \\
NSFNet      & 14  & 0.758 & 0.750 & 0.989 & 0.998 & 0.880 \\
Cost266     & 37  & 0.774 & 0.747 & 0.987 & 0.997 & 0.889 \\
\bottomrule
\end{tabular}
\end{table}

\subsection{RQ8 — GNPy-Validated Generalization}
\label{subsec:rq8}

Finally, we evaluated whether the surrogate trained on analytical GN-model traces generalizes well to physically-validated GNPy simulation outputs, and whether it transfers across topology boundaries without needing to retrain. For each backbone topology (NSFNet, Cost266), we generated 8{,}000 GN-model training traces. We evaluated the surrogate models on three held-out sets, namely, (i)~in-distribution GN-model traces, (ii)~GNPy-simulated traces on the same topology, and (iii)~GNPy-simulated traces on the \emph{other} topology (cross-topology transfer). Across all three trace sets, we used the balanced feasibility splits ($\sim$40\% feasible) by sampling launch powers from $[-18, +4]$\,dBm and higher-order modulation formats (32QAM, 64QAM) at elevated probabilities.

\paragraph{Results.} Table~\ref{tab:rq8_gnpy} shows MAE, RMSE, feasibility accuracy, AUC, and P10--P90 interval coverage across all six chosen evaluation conditions. We found MAE to increase by less than 0.10\,dB when moving from in-distribution GN-model data to GNPy-simulated traces (0.380$\to$0.431\,dB for NSFNet, 0.394$\to$0.431\,dB for Cost266). These findings confirm that the surrogate generalizes from the analytical training simulator to the physically validated GNPy model.

For cross-topology setting, we found MAE to remain within 0.09\,dB of in-distribution performance on Cost266 (0.395\,dB) and increases to 0.472\,dB on the NSFNet$\to$Cost266 transfer -- this is consistent with the slightly wider path-length distribution of Cost266 versus NSFNet. AUC remained at 0.997--0.999 across all six evaluation conditions, meaning that the feasibility rank ordering is robust to both emulator and topology shifts. Finally, we found that the interval coverage degraded from 0.91 to 0.78 on the hardest cross-topology condition (NSFNet$\to$Cost266), which motivates a lightweight post-hoc recalibration as noted in RQ5.

\begin{table}[t]
\centering
\caption{GNPy-validated generalization. Surrogate trained on GN-model traces, evaluated on GNPy-simulated and cross-topology traces.}
\label{tab:rq8_gnpy}
\small
\setlength{\tabcolsep}{3pt}
\begin{tabular}{lrrrrrr}
\toprule
\textbf{Eval set} & \textbf{MAE} & \textbf{RMSE} &
\textbf{Acc} & \textbf{AUC} &
\textbf{Cov\textsubscript{10--90}} & \textbf{Width} \\
\midrule
NSFNet GN (in-dist)   & 0.380 & 0.475 & 0.975 & 0.998 & 0.907 & 1.55\,dB \\
NSFNet GNPy (OOD)     & 0.431 & 0.540 & 0.977 & 0.998 & 0.834 & 1.54\,dB \\
NSFNet$\to$Cost266    & 0.472 & 0.606 & 0.973 & 0.997 & 0.780 & 1.50\,dB \\
\midrule
Cost266 GN (in-dist)  & 0.394 & 0.492 & 0.975 & 0.998 & 0.915 & 1.66\,dB \\
Cost266 GNPy (OOD)    & 0.431 & 0.538 & 0.971 & 0.997 & 0.875 & 1.65\,dB \\
Cost266$\to$NSFNet    & 0.395 & 0.496 & 0.984 & 0.999 & 0.895 & 1.65\,dB \\
\bottomrule
\end{tabular}
\end{table}

\subsection{RQ9 — Inference Latency on Production Network Feature Vectors}

We evaluated whether the surrogate's sub-millisecond latency advantage holds when queried on feature vectors derived from a live production optical network. We used the Alibaba Cloud Transport System dataset~\cite{alibaba2024}, which provides hourly optical performance monitoring telemetry from a production deployment over a 15-day window. The dataset includes per-amplifier gain measurements, OCM per-channel power, vendor NF-gain curves, and pre-FEC BER per transponder.

\paragraph{Setup.} We extracted 11,209 telemetry snapshots from the dataset. For each snapshot, we derived the surrogate input feature vector: span count, span length (from amplifier gain and fiber attenuation coefficient), launch power (from OCM channel power), noise figure (interpolated from vendor NF-gain curves), active channel count, baud rate, and required OSNR (from the dataset's BER-GOSNR transponder curves). We trained a surrogate on GN-model simulator traces matched to the network's operating conditions: launch power $[-5, +8]$~dBm, NF $[4.5, 9.0]$~dB, channel count $[2, 15]$, and span length $[70, 140]$~km. The simulator-only training pipeline from Section~5 was unchanged.

\paragraph{Latency results.} Table~\ref{tab:latency_alibaba} reports single-query and candidate-set latency on the Alibaba feature vectors alongside the GNPy reference. The surrogate achieves $0.41 \pm 0.29$~ms single-query latency on production network inputs — a $6{,}275\times$ speedup over GNPy. A 1{,}000-path candidate set is evaluated in 1.23~ms, compared to 43.2 minutes for GNPy. These results are consistent with the simulator-derived latency in RQ2 (Tab.~\ref{tab:rq2_latency}). The latency advantage holds regardless of whether the input features originate from a simulator or a live network.

\begin{figure}[t]
\centering
\includegraphics[width=0.55\columnwidth]{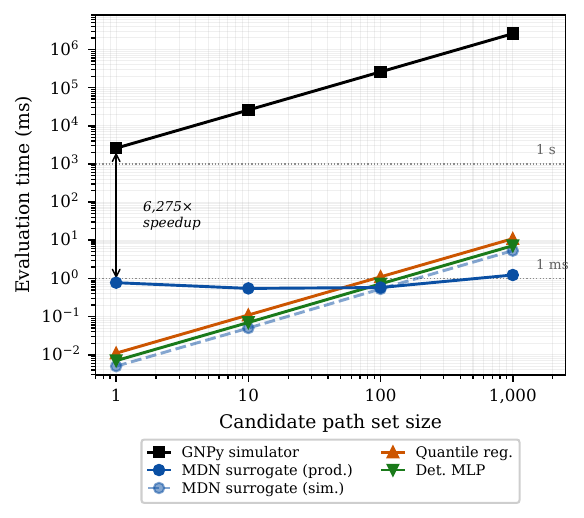}
\caption{Candidate-set evaluation latency for GNPy, MDN surrogate (simulator inputs and Alibaba production inputs), quantile regression, and deterministic MLP. The MDN surrogate achieves a $6{,}275\times$ speedup over GNPy on production feature vectors, with sub-millisecond latency even for 1{,}000-path candidate sets.}
\label{fig:rq9_latency}
\end{figure}

\paragraph{Distributional scope.} The Alibaba network operates in a narrow high-margin regime: Q-margin mean $= 7.27$~dB, std $= 0.45$~dB, and feasibility rate $= 1.000$. This does not exercise the surrogate's discriminative capability near the feasibility boundary. Absolute Q-margin MAE on live hardware was 3.44~dB. The same surrogate achieves MAE $= 0.43$~dB on held-out simulator traces from the same operating regime, which confirms that the degradation is driven by the narrow distributional scope of the dataset rather than a modeling limitation. A meaningful margin accuracy evaluation on live hardware requires traces from a network operating near the provisioning boundary. We identify this as a direction for future work.
\begin{table}[t]
\centering
\caption{Candidate-set evaluation latency on Alibaba production network feature vectors. GNPy latency from Section~\ref{subsec:rq2}.}
\label{tab:latency_alibaba}
\small
\setlength{\tabcolsep}{4pt}
\begin{tabular}{lrrrr}
\toprule
Method & 1 path & 10 paths & 100 paths & 1{,}000 paths \\
\midrule
GNPy simulator   & 2{,}592~ms & 25.9~s  & 259~s   & 43.2~min \\
MDN surrogate    & 0.77~ms    & 0.55~ms & 0.58~ms & 1.23~ms  \\
\bottomrule
\end{tabular}
\end{table}

\section{Conclusion and Limitations}\label{sec:conclusion}

In this paper, we presented a generative surrogate for the optical layer as a first case study in fast path feasibility estimation at runtime. The surrogate is trained offline once on simulator-generated traces. The surrogate is queried online by the controller, which exposes the optical-layer behavior as a fast, probabilistic query engine. This serves as a replacement for offline simulation calls or deterministic feasibility rules that are commonly used.

We evaluated the surrogate along nine research questions. Across nine research questions, the surrogate was found to match deterministic predictor accuracy on Q-margin estimation (MAE\,=\,0.787\,dB). Single-path query latency was reduced from 2.592\,s to 0.005\,ms relative to the GNPy simulator. The calibrated MDN matched quantile regression on interval coverage while supporting arbitrary violation thresholds at runtime without retraining. The generative risk-aware policy achieved the lowest blocking probability and the highest spectrum utilization among all four evaluated policies. We have shown that feasibility rank ordering partially survived cross-emulator shift to Mininet-Optical (AUC\,0.667). We have shown that MDN proactive controller eliminated connection failures by triggering rerouting 19 steps earlier than the reactive MLP baseline. Surrogate MAE was found to remain flat at 0.72--0.77\,dB across topologies ranging from synthetic benchmarks to real 37-node backbones. Finally, GNPy-validated generalization showed that MAE degraded by less than 0.10\,dB and AUC remained at 0.997--0.999 under both out-of-distribution and cross-topology transfer conditions. Finally, inference latency on 11{,}209 live production network snapshots held the sub-millisecond advantage, which is $6{,}275\times$ speedup over the GNPy simulator regardless of whether input features originate from a simulator or a live network.

Our work has several limitations as follows. First, our study entirely relies on simulator-driven traces and no live optical hardware was used. Second, the controller integration covers path-selection and first-fit spectrum assignment only. Joint routing and spectrum optimization under dynamic traffic is not addressed in this work. Third, the dynamic reconfiguration experiment in RQ6 uses a controlled degradation model with a fixed drift rate. Measured amplifier aging traces from live deployments are not used. Fourth, the MDN calibration results indicate that temperature scaling or conformal corrections will likely be needed before the raw interval estimates can be used as confidence bounds by the controller.

Fifth, the surrogate depends on the input features being up to date or current. In practice, the span parameters, amplifier characteristics, and spectrum occupancy need to be refreshed from the network telemetry which change over time. We have not studied how much the predicted Q-margin distribution changes when these telemetry inputs contain measurement errors. Sixth, we evaluated interval coverage only at fixed quantiles. We have not directly checked whether the violation probabilities used by the controller are calibrated, i.e., whether a path with a predicted 10\% violation probability actually violates the threshold 10\% of the time. To account for some of these limitations, we will investigate the surrogate with live hardware traces as part of our future work. We will also investigate the joint RSA optimization and domain-adaptive recalibration for cross-emulator deployment scenarios.

\bibliographystyle{ACM-Reference-Format}
\bibliography{references}

\appendix

\section{Implementation Details}\label{app:implementation}

\subsection{Dataset and Features}
The simulator-only dataset contains 10{,}000 candidate lightpaths sampled across two topology regimes. Each candidate path is represented using controller-visible features, namely, topology identifier, source and destination node identifiers, traffic demand, path length, hop count, span count, launch power, baud rate, modulation format, load factor, and the modulation-dependent required OSNR threshold. The target outputs are OSNR, Q-margin, and feasibility.

We intentionally constructed the dataset to include both feasible and infeasible paths. In the generated dataset, 45.75\% of examples are feasible and 54.25\% are infeasible. This balance avoids a degenerate feasibility-classification problem and provides enough low-margin examples near the decision boundary.

\subsection{Model Configurations}
The deterministic baseline is a multilayer perceptron (MLP) with two hidden layers of sizes 64 and 32. We train separate MLP models for OSNR regression, Q-margin regression, and feasibility classification. Numerical features are standardized and categorical features are one-hot encoded.

The quantile-regression baseline uses gradient-boosted quantile regressors trained for the 0.10, 0.50, and 0.90 quantiles of Q-margin. We use the P10--P90 interval to evaluate empirical interval coverage and calibration error.

The generative surrogate is implemented as a mixture density network (MDN). The MDN maps candidate-path features to a mixture distribution over Q-margin:
\[
p(y \mid x) = \sum_{k=1}^{K} \pi_k(x)\mathcal{N}(y;\mu_k(x),\sigma_k^2(x)),
\]
where \(x\) denotes the candidate-path features and \(y\) denotes Q-margin. Here, \(K=5\). From this distribution, the controller estimates the expected Q-margin, feasibility probability, and threshold-violation probability.

\subsection{Uncertainty Calibration}
We sample the MDN distribution to estimate P10, P50, and P90 values for interval estimation. We apply a simple post-hoc temperature calibration to the sampled standard deviations before computing the intervals. We must note that the calibrated MDN is used only for uncertainty-interval evaluation. For controller decision latency, we use the analytic MDN violation-probability computation rather than repeated sampling.

\subsection{GNPy Runtime Measurement}
We measured the runtime of a GNPy transmission example on the local machine to estimate the simulator-call latency. The measured software runtime was found to be 2.592 seconds for a single path query. We must note that this is distinct from the physical propagation latency that GNPy reported for the optical path. We use this measured software runtime as the simulator-call cost in the candidate-set latency comparison in Tab.~\ref{tab:rq2_latency}.

\section{RSA-lite Controller Details}\label{app:rsa}
The RSA-lite controller evaluates a set of candidate paths for each connection request. Each policy ranks the candidate paths according to its own rule -- following which, the controller attempts first-fit contiguous spectrum assignment on every link of the selected path. A request is blocked if no candidate path satisfies the policy constraint or if the chosen path cannot support a contiguous slot block.

The four evaluated policies are as follows. (i) \textbf{Shortest feasible + FF} selects the shortest simulator-feasible path and then applies first-fit spectrum assignment. (ii) \textbf{Deterministic best-margin + FF} selects the path with the highest deterministic predicted Q-margin among paths predicted feasible by the MLP. (iii) \textbf{Conservative margin + FF} selects a path only if its predicted Q-margin exceeds a fixed safety margin. (iv) \textbf{Generative risk-aware + FF} selects the path with the lowest MDN-estimated violation probability, subject to the selected risk budget.

We map the traffic demand to a required number of contiguous spectrum slots. The first-fit allocator searches for the first available contiguous slot block that is free across all links of the selected path. Spectrum utilization is computed as the fraction of occupied slots across all modeled links. The fragmentation proxy is computed as the fraction of free slots that belong to small free-slot blocks.

\section{Additional Generative Baseline}\label{app:vae}
We also evaluated a conditional variational autoencoder (VAE) as an alternative generative surrogate to justify the MDN model choice. In this first implementation, the VAE was found to be weaker than the MDN for scalar Q-margin distribution learning -- the VAE produced higher Q-margin error and weaker feasibility recall compared to the MDN. Following this, we use the MDN as the primary generative surrogate in the main evaluation.

This result suggests that for low-dimensional physical-layer targets such as Q-margin, mixture density models may be a better first instantiation than latent-variable reconstruction models. VAEs may become more useful in future work where the surrogate is required to generate higher-dimensional physical-layer states rather than a scalar path-quality metric.

\section{Use of Generative AI}\label{app:ai}
 All technical content, experimental design, results, analysis, and scientific claims are solely the work of the authors. No references or citations were generated using generative AI. Generative AI was used to assist with debugging portions of the code and for conciseness in selected passages -- these specific items were thoroughly verified by the author.

\end{document}